\documentclass{article}
\usepackage{spconf,amsmath,graphicx}

\usepackage{enumitem}
\setlist{nosep, leftmargin=14pt}

\usepackage{mwe} 
\usepackage{multirow}
\usepackage{amssymb}
\usepackage{amsmath}

\title{DINO-AD: Unsupervised Anomaly Detection with Frozen DINO-V3 Features}
\name{Jiayu Huo$^{1,\dagger}$ \qquad Jingyuan Hong$^{2,\dagger}$ \qquad Liyun Chen$^{3}$\thanks{${\dagger}$ Equal Contribution \\ Corresponding author: j.huo@ic.ac.uk}}

\address{
$^{1}$ National Heart and Lung Institute, Imperial College London, London, UK \\
$^{2}$ Department of Digital Twins for Healthcare, BMEIS, King’s College London, London, UK \\
$^{3}$ SonoScape Medical Corp.}
\begin{document}
%
\maketitle
\begin{abstract}
Unsupervised anomaly detection (AD) in medical images aims to identify abnormal regions without relying on pixel-level annotations, which is crucial for scalable and label-efficient diagnostic systems. In this paper, we propose a novel anomaly detection framework based on DINO-V3 representations, termed DINO-AD, which leverages self-supervised visual features for precise and interpretable anomaly localization. Specifically, we introduce an embedding similarity matching strategy to select a semantically aligned support image and a foreground-aware K-means clustering module to model the distribution of normal features. Anomaly maps are then computed by comparing the query features with clustered normal embeddings through cosine similarity. Experimental results on both the Brain and Liver datasets demonstrate that our method achieves superior quantitative performance compared with state-of-the-art approaches, achieving AUROC scores of up to 98.71. Qualitative results further confirm that our framework produces clearer and more accurate anomaly localization. Extensive ablation studies validate the effectiveness of each proposed component, highlighting the robustness and generalizability of our approach.
\end{abstract}

\begin{figure*}[!htbp]
\centering
\includegraphics[width=\linewidth]{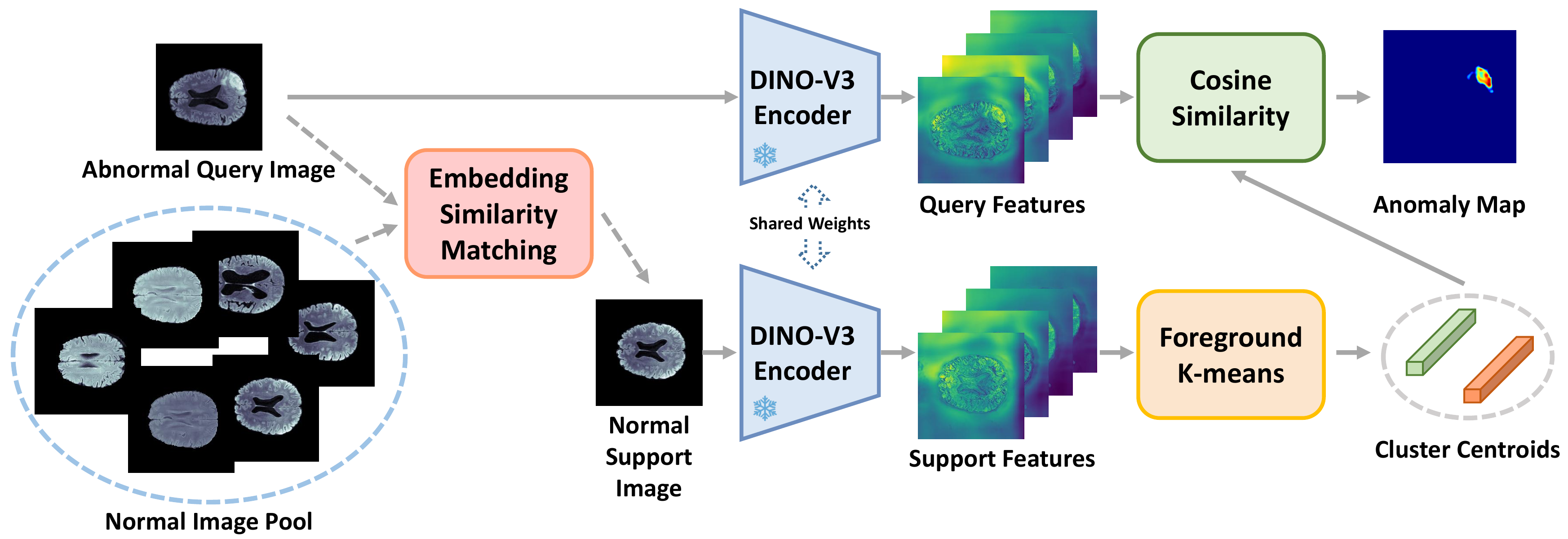}
\caption{Overview of DINO-AD framework. Our pipeline is training-free, which allows it to generalize across diverse datasets and anomaly types without requiring labeled data or prior model adaptation.}
\label{fig:main_fig}
\end{figure*}

\section{Introduction}
\label{sec:introduction}
Anomaly detection (AD) in medical images plays a crucial role in computer-aided diagnosis by identifying abnormal regions that may indicate pathological changes. Traditional supervised approaches rely heavily on large-scale annotated datasets, which are difficult and costly to obtain, especially in medical domains where expert labeling is required. To overcome this limitation, unsupervised anomaly detection (UAD) has recently gained increasing attention. UAD methods aim to model the distribution of normal data and detect deviations from this distribution as anomalies, making them more scalable and practical for clinical applications where abnormal samples are scarce or unavailable.

Many recent studies have proposed training-based anomaly detection frameworks that rely on reconstruction or feature distillation mechanisms, such as DRAEM~\cite{zavrtanik2021draem}, UTRAD~\cite{chen2022utrad}, RD4AD~\cite{deng2022anomaly}, STFPM~\cite{wang2021student}, and PaDiM~\cite{defard2021padim}. While these methods have achieved promising results, they typically require an additional training stage to adapt models to specific datasets. Such dependency often limits their generalization ability across different imaging modalities and anatomical structures. Moreover, the training process introduces extra computational cost and parameter sensitivity, which may hinder their deployment in real-world medical scenarios.

In contrast, several training-free approaches have been proposed to alleviate the need for dataset-specific fine-tuning, such as PatchCore~\cite{roth2022towards}, AnomalyDINO~\cite{damm2025anomalydino}, and DINO-DPMM~\cite{schulthess2025anomaly}. These methods leverage pre-trained vision transformers or feature memory banks to perform anomaly detection directly in the embedding space. Although they demonstrate strong generalization and efficiency, they still suffer from limitations in accurately capturing semantic correspondence between normal and query images. In particular, the lack of foreground awareness and feature clustering often leads to incomplete localization or false activation in background regions, reducing detection precision.

To address these challenges, we propose a training-free unsupervised anomaly detection framework based on DINO-V3 representations, named DINO-AD. The method incorporates an embedding similarity matching strategy to select a semantically aligned normal support image and a foreground-aware K-means clustering module to model diverse normal feature prototypes. By combining these components, our approach generates more accurate and spatially consistent anomaly maps without any additional training. Experiments on Brain and Liver datasets show that DINO-AD achieves state-of-the-art performance in both quantitative and qualitative evaluations, demonstrating its effectiveness and strong generalization ability for medical anomaly detection.

\section{Methods}
\label{sec:methods}
The DINO-AD architecture is shown in Figure~\ref{fig:main_fig}. We first use the embedding similarity matching algorithm (Section~\ref{subsec:embedding_similarity_matching}) to find the normal support image with the closest appearance based on the abnormal query image. Then the query and support images go through the DINO-V3 encoder to obtain query and support features. We further use the K-means algorithm to compute the cluster centroids in the foreground area (Section~\ref{subsec:foreground_kmeans_cluster}). Finally, the anomaly map is computed according to the cosine similarity between query features and cluster centroids (Section~\ref{subsec:computation_of_anomaly_maps}). We froze the DINO-V3 weights for feature extraction, which makes the pipeline training-free.

\subsection{Embedding Similarity Matching}
\label{subsec:embedding_similarity_matching}
We first feed both the abnormal query image and all normal images into the DINO-V3 model and extract the one-dimensional feature vector from the final layer as the embedding representation. Then, we compute the cosine similarity between the embedding of the abnormal query image and each of the normal image embeddings. Finally, we select the normal image with the highest similarity score as the support image.

\subsection{Foreground K-means Cluster}
\label{subsec:foreground_kmeans_cluster}
We first obtain the foreground mask of the support image by locating all pixels whose values are non-zero and setting them to 1. Then, a morphological closing operation is applied to remove small holes inside the mask. After obtaining the foreground mask, we use DINO-V3 to extract intermediate features of the support image and perform K-means clustering within the foreground region to extract a set of representative cluster center features, which are later used for anomaly map computation.

Formally, given a set of feature vectors \(\{\mathbf{x}_1, \mathbf{x}_2, \ldots, \mathbf{x}_N\}\) within the foreground region, K-means aims to partition them into \(K\) clusters \(\{C_1, C_2, \ldots, C_K\}\) by minimizing the following objective function:

\begin{equation}
\mathcal{L} = \sum_{k=1}^{K} \sum_{\mathbf{x}_i \in C_k} \|\mathbf{x}_i - \boldsymbol{\mu}_k\|_2^2
\end{equation}

where \(\boldsymbol{\mu}_k\) denotes the centroid of cluster \(C_k\). The algorithm iteratively alternates between assigning each feature vector \(\mathbf{x}_i\) to its nearest cluster according to
\begin{equation}
C_k = \{\mathbf{x}_i : \|\mathbf{x}_i - \boldsymbol{\mu}_k\|_2^2 \le \|\mathbf{x}_i - \boldsymbol{\mu}_j\|_2^2, \, \forall j \},
\end{equation}
and updating each centroid as
\begin{equation}
\boldsymbol{\mu}_k = \frac{1}{|C_k|} \sum_{\mathbf{x}_i \in C_k} \mathbf{x}_i.
\end{equation}
These two steps are repeated until convergence, i.e., when the centroids no longer change significantly. The final set of centroids \(\{\boldsymbol{\mu}_1, \ldots, \boldsymbol{\mu}_K\}\) represents the cluster center features used in subsequent anomaly map computation.

\subsection{Computation of Anomaly Maps}
\label{subsec:computation_of_anomaly_maps}
We compute the anomaly map using the intermediate-layer features of the query image extracted from the DINO-V3 model and the cluster centroids obtained from the foreground K-means clustering. Specifically, we first calculate the cosine similarity between each query feature vector and all cluster centroids, and then take the average over all centroids to obtain a cosine similarity map. Let \(\mathbf{f}_p\) denote the DINO-V3 feature vector at patch location \(p\), and let \(\{\boldsymbol{\mu}_1, \boldsymbol{\mu}_2, \ldots, \boldsymbol{\mu}_K\}\) be the set of cluster centroids obtained from the foreground region. The cosine similarity between \(\mathbf{f}_p\) and each centroid \(\boldsymbol{\mu}_k\) is defined as

\begin{equation}
s_{p,k} = \frac{\mathbf{f}_p \cdot \boldsymbol{\mu}_k}{\|\mathbf{f}_p\|_2 \, \|\boldsymbol{\mu}_k\|_2}.
\end{equation}

The mean cosine similarity for patch \(p\) is then given by
\begin{equation}
\bar{s}_p = \frac{1}{K} \sum_{k=1}^{K} s_{p,k}.
\end{equation}

The resulting map is subsequently normalized between 0 and 1. Since the DINO-V3 features are extracted from image patches, we compute the patch anomaly map $A_p$ by taking one minus the normalized cosine similarity map: 
\begin{equation}
A_p = 1 - \text{Norm}(\bar{s}_p),
\end{equation}
where \(\text{Norm}(\cdot)\) denotes min-max normalization. The final anomaly map $A$ is obtained by combining all patch-level scores $A_p$ and is resized to the original image resolution using bilinear interpolation.

\section{Experimental Results}
\label{sec:results}

\subsection{Datasets}
\label{subsec:datasets_and_implementation}
We evaluated the performance of DINO-V3 on the brain and liver anomaly segmentation datasets from the BMAD benchmark~\cite{bao2024bmad}. The brain dataset was obtained from the BraTS 2021 dataset. We use 7500 normal training images and the normal image pool, and 3075 abnormal test images with anomaly masks to evaluate the performance of DINO-AD. The liver dataset was constructed from the BTCV and LiTs datasets. The training set contains 1542 normal images, and the test set contains 660 abnormal images with masks.

\subsection{Comparison Methods and Evaluation Metrics}
\label{subsec:comparison_methods_and_evaluation_metrics}
We compare DINO-AD to the following anomaly detection methods: DREAM~\cite{zavrtanik2021draem}, UTRAD~\cite{chen2022utrad}, RD4AD~\cite{deng2022anomaly}, STFPM~\cite{wang2021student}, PaDiM~\cite{defard2021padim}, PatchCore~\cite{roth2022towards}, AnomalyDINO~\cite{damm2025anomalydino}, and DINO-DPMM~\cite{schulthess2025anomaly}. Note that DREAM, UTRAD, RD4AD, STFPM, and PaDiM need a GPU to train their models. PatchCore, AnomalyDINO, and DINO-DPMM are training-free methods, but they need to maintain a memory bank or fit a statistics model (i.e., Dirichlet Process Mixture model (DPMM)) to generate the anomaly map. We use the area under the receiver operator characteristic curve (AUROC) and the area under the precision-recall curve (AUPRC) at the pixel level to evaluate the performance of DINO-AD and its counterparts.

\begin{table}[!htbp]
\caption{Quantitative performances of DINO-AD and the comparison methods. The best and the second best results are bold and underlined, respectively.}
\label{tab:main_result}
\centering
\fontsize{9}{11}\selectfont
\begin{tabular}{l|cc|cc}
\hline
\multirow{2}*{Model} &\multicolumn{2}{c|}{Brain} &\multicolumn{2}{c}{Liver} \\
&AUROC &AUPRC &AUROC &AUPRC \\
\hline
DRAEM       &92.70 &38.66 &93.82 &16.01 \\
UTRAD       &92.81 &24.63 &84.28 &1.25 \\
RD4AD       &97.81 &52.11 &94.35 &3.48 \\
STFPM       &96.64 &45.45 &94.70 &3.20 \\
PaDiM       &85.77 &8.21  &89.77 &1.66 \\
PatchCore   &\textbf{98.34} &\underline{67.37} &96.67 &5.66 \\
AnomalyDINO &97.71 &58.69 &\underline{97.24} &\underline{17.91} \\
DINO-DPMM   &96.21 &43.43 &95.46 &9.02 \\
Ours        &\underline{98.22} &\textbf{70.52} &\textbf{98.71} &\textbf{21.42}\\
\hline
\end{tabular}
\end{table}

\begin{figure}[!t]
\centering
\includegraphics[width=0.98\linewidth]{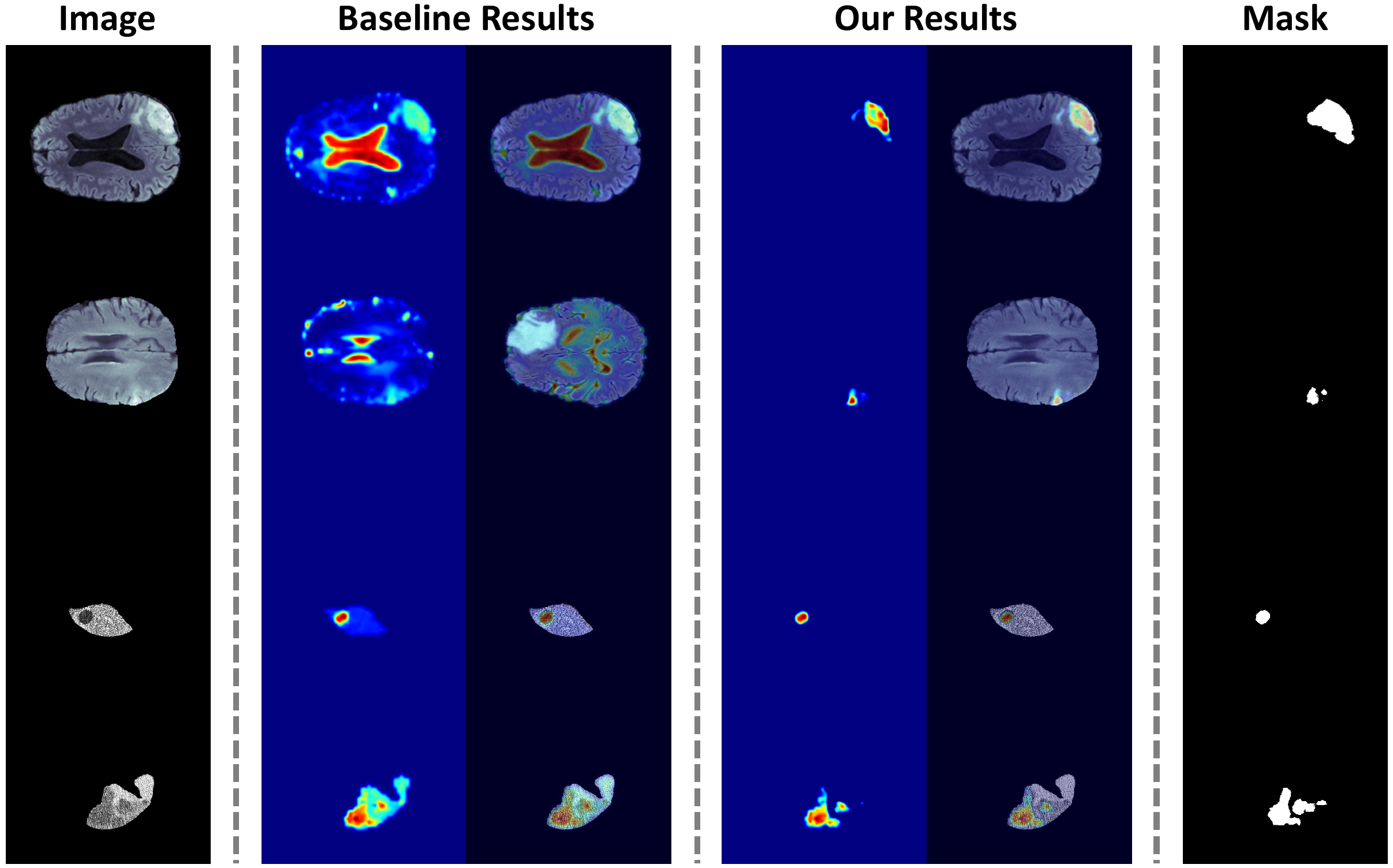}
\caption{Qualitative results of the generated anomaly maps. Here we use the AnomalyDINO as the baseline model.}
\label{fig:visualization_results}
\end{figure}

\subsection{Quantitative and Qualitative Results}
\label{subsec:quantitative_and_qualitative_results}
Table~\ref{tab:main_result} presents the quantitative performance comparison between our proposed method (denoted as Ours) and several state-of-the-art approaches on both the Brain and Liver datasets. Overall, our method achieves the best or second-best results across all metrics. Specifically, for the Brain dataset, our approach attains an AUROC of 98.22 and an AUPRC of 70.52, outperforming most competing methods and slightly trailing PatchCore in AUROC by only 0.12. On the Liver dataset, our model achieves the highest AUROC and AUPRC values of 98.71 and 21.42, respectively, showing a marked improvement over previous methods, including AnomalyDINO and PatchCore. These results demonstrate the robustness and superior anomaly detection capability of our proposed framework in both subtle and complex scenarios.

Figure~\ref{fig:visualization_results} illustrates the qualitative comparison of the generated anomaly maps, where AnomalyDINO is used as the baseline. As shown in the visualization, our method produces anomaly maps with more accurate localization and clearer boundary delineation of the abnormal regions. In contrast, the baseline often yields coarse or incomplete activation in certain areas. The improved visual quality can be attributed to our embedding similarity matching and foreground-aware K-means clustering, which enhance the discriminative power of feature representations. Overall, the qualitative results confirm that our approach provides more precise anomaly localization compared to existing methods.

\begin{table}[!htbp]
\caption{Quantitative results of different numbers of cluster centroids.}
\label{tab:kmeans_ablation}
\centering
\fontsize{9}{11}\selectfont
\begin{tabular}{c|cc|cc}
\hline
Number of &\multicolumn{2}{c|}{Brain} &\multicolumn{2}{c}{Liver} \\
Clusters &AUROC &AUPRC &AUROC &AUPRC \\
\hline
1  &96.54 &66.41 &95.61 &18.01 \\
2  &98.22 &70.52 &98.71 &21.42 \\
3  &97.67 &69.17 &97.14 &20.87 \\
4  &97.19 &68.45 &96.42 &19.74 \\
\hline
\end{tabular}
\end{table}

\begin{table}[!htbp]
\caption{Quantitative results of different support image selection strategies. ESM is short for embedding similarity matching.}
\label{tab:esm_ablation}
\centering
\fontsize{9}{11}\selectfont
\begin{tabular}{c|cc|cc}
\hline
Support Image &\multicolumn{2}{c|}{Brain} &\multicolumn{2}{c}{Liver} \\
Selection &AUROC &AUPRC &AUROC &AUPRC \\
\hline
Random  &84.21 &55.14 &80.19 &9.23 \\
ESM     &98.22 &70.52 &98.71 &21.42 \\

\hline
\end{tabular}
\end{table}

\subsection{Ablation Studies}
\label{subsec:ablation_studies}
Tables~\ref{tab:kmeans_ablation} and~\ref{tab:esm_ablation} present the ablation results evaluating the effects of the number of cluster centroids and the support image selection strategy. As shown, when only one cluster is used, the model performs poorly due to the limited ability to represent the feature diversity within the foreground region. Increasing the number of clusters to two yields the best AUROC and AUPRC values on both Brain and Liver datasets, indicating an optimal balance between feature compactness and diversity. However, using more clusters leads to performance degradation, likely caused by feature over-segmentation and noise from small or redundant clusters. In addition, replacing the embedding similarity matching (ESM) strategy with random support image selection results in a substantial drop in performance, confirming the importance of semantic alignment between the query and normal images. These results collectively demonstrate that both an appropriate number of cluster centroids and an effective support image selection strategy are essential for achieving robust and accurate anomaly detection.

\section{Conclusion}
\label{sec:conclusion}
In this paper, we propose a novel unsupervised anomaly detection framework based on DINO-V3 representations, incorporating embedding similarity matching and foreground-aware K-means clustering. The proposed method effectively leverages semantic-rich self-supervised features and foreground structure information to enhance anomaly localization and discrimination. Extensive experiments on the Brain and Liver datasets demonstrate that our approach outperforms existing state-of-the-art methods in both quantitative and qualitative evaluations. Furthermore, ablation studies validate the effectiveness of each component, including the number of cluster centroids and the support image selection strategy. In the future, we plan to extend this framework to more complex medical imaging modalities and explore its potential for real-time clinical applications. Moreover, we aim to investigate adaptive feature selection and cross-domain generalization strategies to further enhance the robustness and scalability of our anomaly detection framework.


\bibliographystyle{IEEEbib}
\bibliography{strings,refs}

\end{document}